\documentclass[11pt,a4paper]{article}
\usepackage[utf8]{inputenc}
\usepackage[T1]{fontenc}
\usepackage[english]{babel}
\usepackage{here}
\usepackage{graphicx}
\usepackage{subcaption}
\usepackage{eurosym}
\usepackage{amsmath}
\usepackage{amssymb}
\usepackage{latexsym}
\usepackage{xcolor}
\usepackage{mathrsfs}
\usepackage{caption}
\usepackage{braket}
\usepackage{float}
\usepackage[compact]{titlesec}

\usepackage{hyperref}
\usepackage[style=numeric,sorting=none,url=true,doi=true,backend=bibtex]{biblatex}
\usepackage{cleveref}
\usepackage[switch]{lineno}
\usepackage[font=footnotesize,labelfont=bf]{caption}
\usepackage[hmargin=2cm,vmargin=2cm,rmargin=2cm, lmargin=2cm]{geometry}
\hypersetup{colorlinks=true,citecolor=red,filecolor=blue,linkcolor=blue,urlcolor=blue}
\DeclareUnicodeCharacter{FB01}{fi}
\DeclareUnicodeCharacter{200B}{}

\def\keywordname{{\bfseries \emph{Keywords}}}%
\def\keywords#1{\par\addvspace\medskipamount{\rightskip=0pt plus1cm
\def\and{\ifhmode\unskip\nobreak\fi\ $\cdot$
}\noindent\keywordname\enspace\ignorespaces#1\par}}
\begin{document}
\renewcommand{\figurename}{\textbf{Fig.}}
\renewcommand{\tablename}{\textbf{Table}}
\newcommand{\un}{\mathds{1}}
\newcommand{\1}{\mathbbm{1}}
\newcommand{\vect}[1]{\boldsymbol{#1}}
\newcommand{\era}{\end{array}}
\newcommand{\beq}{\begin{equation}}
\newcommand{\eeq}{\end{equation}}
\newcommand{\beqar}{\begin{eqnarray}}
\newcommand{\eeqar}{\end{eqnarray}}
\newcommand{\lb}{\label}
\thispagestyle{empty}
\baselineskip=18pt
\medskip
\begin{center}
~~~~~~~~~~~~~~~
\\
\vspace{2cm}
\noindent { {\textbf{ Quantum Entanglement in the Dirac Field Quantization around Charged Black Holes. }}}\\

%%%%%%%%%%%%%%%%%%%%%%%%%%%%%%%%%%%%%%%%%%%%%%%%%%%%%%%%%%%%%%%%%%%%%%%%%%%%%%%%%%%%%%%%%%%%%%%%%%%%%%%%%%%%%%%%%%%%%
\vspace{0.7cm}
\noindent
\vspace{0.5cm}
{\small Abdessamie Chhieb $^{a,b,}$}{\footnote{E-mail:
\textsf{\href{mailto:chhiebabdessamie@gmail.com }{chhiebabdessamie@gmail.com  }}}},  {\small Chaimae Banouni $^{a,b,}$}{\footnote{E-mail:
\textsf{\href{mailto:banounichaymae1@gmail.com }{banounichaymae1@gmail.com  }}}}, {\small Saliha Abdessamie $^{a,b,}$}{\footnote{E-mail:
\textsf{\href{mailto:salihaabdessamia67@gmail.com }{salihaabdessamia67@gmail.com}}}}, 
	{\small Mohamed Ouchrif $^{a,b,}$}\footnote{E-mail: \textsf{\href{mailto:Mohamed.Ouchrif@cern.ch}{mohamed.ouchrif@cern.ch}}},

\noindent $^{a}${{\footnotesize   Laboratory of Theoretical Physics, Particles, Modeling and Energies, Faculty of Sciences,\\ Mohamed First University, Oujda,  Morocco}}\\[0.5em]
\noindent $^{b}${{\footnotesize  National Institute For Particle Physics and Applications (NIPPA), Oujda,  Morocco}}\\[0.5em]

\end{center}
\vspace{0.5cm}
\begin{abstract}
We investigate the quantum entanglement properties of the Dirac field near a charged Reissner--Nordström black hole, incorporating the effects of Hawking radiation within the framework of quantum field theory in curved spacetime. Using concurrence \( C \) and Bures distance \( B \) as measures of entanglement, we analyze how quantum correlations evolve with respect to the electric charge \( Q \) of the black hole, the frequency \( \omega \) of fermionic modes, and the initial entanglement angle \( \theta \). Our results show that the electric charge \( Q \) enhances decoherence inside the event horizon while, counterintuitively, temporarily increasing accessible entanglement outside. The Hawking effect induces an apparent loss of entanglement for an external observer, due to correlation transfer to inaccessible regions. High-frequency modes \( \omega \) exhibit greater resilience to gravitational effects, maintaining robust correlations near the horizon. These findings highlight the redistribution of entanglement in a multipartite system in curved spacetime, with significant implications for quantum information in relativistic and gravitational contexts.

\end{abstract}
\vspace{0.5cm}
\newpage
\section*{Introduction }

Quantum entanglement is one of the most remarkable and fundamental features of quantum mechanics \cite{horodecki2009quantum, brunner2014bell}. It refers to correlations between quantum systems that cannot be described classically, making the combined system’s state inseparable into independent parts. Since its discovery, entanglement has become the cornerstone of many quantum technologies. including quantum computing, secure quantum communication, quantum teleportation, and quantum metrology \cite{horodecki2009quantum, nielsen2010quantum,bennett1993teleporting, giovannetti2011advances}. Studying entanglement in relativistic and gravitational contexts is essential to deepen our understanding of the quantum world and to explore the limits of quantum technologies \cite{fuentes2010entanglement}. A particularly fascinating domain is the behavior of entanglement in curved spacetime, where gravity influences the quantum fields. Quantum field theory in curved spacetime, developed notably by Birrell and Davies \cite{birrell1984quantum}, provides the theoretical framework to analyze such phenomena. Near black holes, Hawking radiation \cite{hawking1975particle} predicts the spontaneous creation of particle-antiparticle pairs near the event horizon, altering the quantum vacuum state and affecting entanglement properties. Entanglement in such extreme environments is complex because observers located in different regions of spacetime (e.g., inside versus outside the horizon) cannot access the same information \cite{zhang2025entanglement}. This inaccessibility leads to the redistribution of entanglement among multipartite systems, an idea formalized in studies using Bogoliubov transformations and entanglement measures \cite{AlsingMilburn2003, FuentesSchuller2005, Montero2011}. For example, a maximally entangled state shared between two observers may appear degraded for one observer outside the black hole because part of the entanglement is transferred to inaccessible regions \cite{FuentesSchuller2005, Montero2011}.

Reissner–Nordström black holes present a richer horizon structure than neutral Schwarzschild black holes, with both outer and inner horizons. This structure significantly impacts the quantum field dynamics and entanglement evolution \cite{chandrasekhar1983mathematical, birrell1984quantum}. The electric charge \( Q \) modulates the Hawking radiation spectrum and the location of horizons, influencing how entanglement distributes throughout spacetime.Studying Dirac fields, which describe fermions, is motivated by their intrinsic spin-\(1/2\) nature and Fermi-Dirac statistics, differing from simpler scalar fields \cite{hawking1975particle, kanti2002calculable}. These properties affect the entanglement structure and the response to gravitational effects \cite{neznamov2018stationary}. Thus, analyzing entanglement dynamics of Dirac fields near charged black holes provides a more realistic understanding of quantum information in curved spacetime.

To quantify entanglement, we use concurrence \( C \), well-suited for bipartite two-qubit systems, and the Bures distance \( B \), a geometric measure capturing state closeness robustly even in noisy or open systems \cite{banouni2025unveiling, chhieb2024time, chhieb2024metrological}. By systematically studying the influence of the electric charge \( Q \), fermionic mode frequency \( \omega \), and initial entanglement angle \( \theta \), we provide a detailed characterization of entanglement dynamics near charged black holes.

In this context, it becomes essential to understand how the presence of a strong gravitational field or an event horizon influences the dynamics of quantum correlations and associated resources such as entanglement and coherence. Recent studies have shown that gravity and acceleration can lead to significant degradation of quantum correlations between spatially separated systems, thereby affecting quantum information processing in relativistic regimes~\cite{Peres2004Quantum, alsing2003teleportation, Rideout2012Fundamental}. In particular, the presence of an event horizon (as in Reissner–Nordström black holes) alters the structure of the quantum vacuum and gives rise to thermal radiation effects such as Hawking radiation, which perturb entangled states~\cite{Hawking1975, BirrellDavies, fuentes2010diracEntanglement}. It is thus highly motivating to explore the evolution of quantum resources in such curved spacetime geometries, relying on tools from quantum field theory and relativistic quantum information.

Our results show that the electoral charge \( Q \) enhances decoherence inside the horizon, making correlations fragile in that region. However, counterintuitively, it can temporarily increase accessible entanglement outside the horizon due to multipartite redistribution . High-frequency modes \( \omega \) demonstrate greater robustness to gravitational effects, preserving entanglement near the horizon. Furthermore, the initial state choice, especially Bell-type states with \( \theta = \pi/4 \), plays a crucial role in entanglement preservation. These findings have profound implications for the black hole information paradox, multipartite quantum systems in gravity, and the design of quantum protocols operating in relativistic regimes. They contribute to the understanding of quantum information’s fate in strong gravitational fields and the development of quantum technologies for extreme environments \cite{alsing2003teleportation, mann2012relativistic}.
The paper is organized as follows. Section \ref{sec1} we present the theoretical framework, including the description of the Dirac field in Reissner–Nordström spacetime and the modeling of Hawking radiation effects . In Section \ref{sec2}, introduces the entanglement measures used in this work, namely concurrence and Bures distance, and explains how they are computed in curved spacetime . Section \ref{sec3} presents the numerical simulations and analyzes the influence of the black hole charge \( Q \), the fermionic mode frequency \( \omega \), and the initial entanglement angle \( \theta \) on quantum correlations both inside and outside the event horizons. Finally, Section \ref{sec4} summarizes the main findings, discusses their implications, and outlines possible directions for future research.
\\
\section{Model}
\subsection{Fields and Hawking Radiation in Reissner--Nordstr\"om Black Holes}
\label{sec1}
The  Reissner--Nordstr\"om (RN) black hole is described by the flowing metric  
\begin{equation}
    ds^2 = -f(r)\,dt^2 + \frac{1}{f(r)}\,dr^2 + r^2(d\theta^2 + \sin^2\theta\,d\phi^2),
    \label{eq:metric}
\end{equation}
with the metric function given by
\begin{equation}
    f(r) = 1 - \frac{2GM}{r} + \frac{GQ^2}{r^2},
    \label{eq:f_of_r}
\end{equation}
where $M$ is the mass and $Q$ the electric charge of the black hole, and $G$ is Newton's gravitational constant. In the limit $Q = 0$, this reduces to the Schwarzschild metric~\cite{Carter1966,Bardeen1973}.\\
To explore quantum effects such as Hawking radiation in this background, we study the dynamics of spin-$\frac{1}{2}$ fermionic fields, described by the Dirac equation minimally coupled to gravity and a $U(1)$ gauge field. In curved spacetime, the Dirac equation takes the form~\cite{Jing2004}:
\begin{equation}
    \left[\gamma^a e_a^\mu (\partial_\mu + \Gamma_\mu - i Q A_\mu)\right]\psi = 0,
    \label{eq:dirac}
\end{equation}
where $\gamma^a$ are the flat-space Dirac matrices, $e_a^\mu$ are the inverse vierbein fields satisfying $g_{\mu\nu} = e_\mu^a e_\nu^b \eta_{ab}$, and $\Gamma_\mu$ is the spin connection. The electromagnetic potential for the RN black hole is given by $A_\mu dx^\mu = (-Q/r + \text{const}) dt$. Near the event horizon, it is useful to introduce the tortoise coordinate $r^*$:
\begin{equation}
    \frac{dr^*}{dr} = \frac{1}{f(r)} \quad \Rightarrow \quad r^* = \int \frac{dr}{f(r)},
    \label{eq:tortoise}
\end{equation}
which allows regular null coordinates via $u = t - r^*$. The spinor field solutions then decompose into positive and negative frequency modes:
\begin{equation}
    \psi_k^+ = \xi(r,\theta,\phi)\, e^{-i \omega u}, \quad
    \psi_k^- = \xi(r,\theta,\phi)\, e^{i \omega u},
    \label{eq:spinor_modes}
\end{equation}
where $\omega$ is the mode frequency and $\xi$ encodes angular and radial components. The Hawking temperature is determined by the surface gravity $\kappa$ at the outer horizon $r_+$, where $f(r_+)=0$. It is given by:
\begin{equation}
    T_H = \frac{\kappa}{2\pi} = \frac{1}{4\pi} f'(r_+),
    \label{eq:hawking_temp_general}
\end{equation}
which for the RN metric yields:
\begin{equation}
    T_H = \frac{1}{4\pi r_+} \left(1 - \frac{G Q^2}{r_+^2}\right).
    \label{eq:hawking_temp_rn}
\end{equation}
In the limit $Q \to 0$, the RN black hole becomes Schwarzschild, and the Hawking temperature reduces accordingly. This transition illustrates how electric charge suppresses Hawking radiation by reducing surface gravity~\cite{Hawking1975,HawkingPage1983}.

\subsection{Entanglement Structure in the Presence of Hawking Radiation \label{sec2}}

According to quantum field theory in curved spacetime, the Kruskal vacuum---which describes the full extended black hole---is seen by external observers as an entangled state. Through Bogoliubov transformations, the vacuum decomposes as:
\begin{equation}
    |0\rangle_k = \cos r_\omega\, |0\rangle_I |0\rangle_{II} + \sin r_\omega\, |1\rangle_I |1\rangle_{II},
    \label{eq:kruskal_vacuum}
\end{equation}
where $\cos r_\omega = 1/\sqrt{1 + e^{-\omega/T_H}}$ and $\sin r_\omega = e^{-\omega/(2T_H)}/\sqrt{1 + e^{-\omega/T_H}}$. The states $|0\rangle_I$, $|1\rangle_I$ correspond to the outside region, while $|0\rangle_{II}$, $|1\rangle_{II}$ belong to the interior.
To assess the effect of this structure on quantum correlations, consider the initial bipartite state:
\begin{equation}
    |\Psi\rangle_{AB} = \cos\theta\,|00\rangle + \sin\theta\,|11\rangle,
    \label{eq:bell_state}
\end{equation}
where $\theta$ controls the degree of entanglement. If qubit $B$ approaches the horizon and undergoes mode splitting, the state becomes tripartite:
\begin{equation}
    |\Psi\rangle_{AB_I B_{II}} = \cos\theta\, \cos r_\omega\, |000\rangle + \cos\theta\, \sin r_\omega\, |011\rangle + \sin\theta\, |110\rangle,
    \label{eq:tripartite}
\end{equation}
with $B_I$ and $B_{II}$ denoting modes outside and inside the horizon, respectively. Tracing over $B_{II}$ gives the reduced state:
\begin{equation}
\varrho_{AB_I} = 
\begin{pmatrix}
    \cos^2\theta\, \cos^2 r_\omega & 0 & 0 & \cos\theta\, \sin\theta\, \cos r_\omega \\
    0 & \cos^2\theta\, \sin^2 r_\omega & 0 & 0 \\
    0 & 0 & 0 & 0 \\
    \cos\theta\, \sin\theta\, \cos r_\omega & 0 & 0 & \sin^2\theta
\end{pmatrix},
\label{eq:rho_ABI}
\end{equation}
while tracing over $B_I$ yields:
\begin{equation}
\varrho_{AB_{II}} = 
\begin{pmatrix}
    \cos^2\theta\, \sin^2 r_\omega & 0 & 0 & \cos\theta\, \sin\theta\, \sin r_\omega \\
    0 & \cos^2\theta\, \cos^2 r_\omega & 0 & 0 \\
    0 & 0 & 0 & 0 \\
    \cos\theta\, \sin\theta\, \sin r_\omega & 0 & 0 & \sin^2\theta
\end{pmatrix}.
\label{eq:rho_ABII}
\end{equation}
These matrices illustrate how Hawking radiation redistributes entanglement across the event horizon, leading to a reduction in observable correlations for external observers while maintaining the overall purity of the total system.
\section{Quantum Resources}
\label{sec3}

\subsection{Concurrence}

Concurrence is a widely used entanglement measure tailored for bipartite systems composed of two qubits. For a given density matrix \(\varrho\), the concurrence \(\mathcal{C}(\varrho)\) is defined as~\cite{wootters1998quantum}:
\begin{equation}
\mathcal{C}(\varrho) = \max \left( 0, \, \sqrt{\lambda_1} - \sqrt{\lambda_2} - \sqrt{\lambda_3} - \sqrt{\lambda_4} \right),
\end{equation}
where \(\lambda_i\) (\(i = 1, \dots, 4\)) are the eigenvalues in decreasing order of the non-Hermitian matrix \(\varrho \, \tilde{\varrho}\), with
\begin{equation}
\tilde{\varrho} = (\sigma_Y \otimes \sigma_Y)\, \varrho^{*} \, (\sigma_Y \otimes \sigma_Y),
\end{equation}
and where \(\sigma_Y\) is the Pauli matrix along the \(Y\)-axis and \(\varrho^{*}\) denotes the complex conjugate of \(\varrho\) in the computational basis. In the context of our model, where the initially entangled state is subject to Hawking-induced mode mixing, the reduced density matrices \(\varrho_{AB_I}\) and \(\varrho_{AB_{II}}\) are of X-type form. For such states, the concurrence simplifies :
\begin{align}
\mathcal{C}_{AB_I} &= \frac{2\cos\theta \sin\theta}{\sqrt{1 + e^{-\omega/T_H}}}, \\
\mathcal{C}_{AB_{II}} &= \frac{2\cos\theta \sin\theta \, e^{-\omega/(2T_H)}}{\sqrt{1 + e^{-\omega/T_H}}}.
\end{align}
where \(\theta\) is the initial entanglement parameter, \(\omega\) is the field frequency, and \(T_H\) is the Hawking temperature. These expressions reflect how the entanglement is redistributed across the event horizon as a result of Hawking radiation.

\subsection{Bures Distance}

The Bures distance provides a geometric measure of distinguishability between quantum states, and can be used to quantify entanglement by evaluating how far a given bipartite state \(\rho_{AB}\) lies from the set \(\mathcal{S}\) of all separable states~\cite{uhlmann1976transition,uhlmann1995geometric}:
\begin{equation}
\mathcal{B}(\rho_{AB}) = \sqrt{2 - 2\sqrt{\mathcal{F}(\rho_{AB})}},
\end{equation}
where \(\mathcal{F}(\rho_{AB})\) is the Uhlmann fidelity between \(\rho_{AB}\) and its closest separable state \(\xi \in \mathcal{S}\). In our setup, we consider the Bures distance between the reduced state \(\varrho_{AB_I}\) (or \(\varrho_{AB_{II}}\)) and the initial pure entangled state \(|\Psi\rangle_{AB} = \cos\theta |00\rangle + \sin\theta |11\rangle\). In this case the fidelity simplifies to:
\begin{align}
\mathcal{F}_{AB_I} &= \cos^4\theta \, \cos^2 r_\omega + \cos^2\theta \sin^2\theta \cos r_\omega + \sin^4\theta, \\
\mathcal{F}_{AB_{II}} &= \cos^4\theta \, \sin^2 r_\omega + \cos^2\theta \sin^2\theta \sin r_\omega + \sin^4\theta,
\end{align}
where the auxiliary functions \(\cos r_\omega\) and \(\sin r_\omega\) are defined by:
\begin{equation}
\cos r_\omega = \frac{1}{\sqrt{1 + e^{-\omega/T_H}}}, \quad 
\sin r_\omega = \frac{e^{-\omega/(2T_H)}}{\sqrt{1 + e^{-\omega/T_H}}}.
\end{equation}

The corresponding Bures distances  are:
\begin{align}
\mathcal{B}_{AB_I} &= \sqrt{2\left(1 - \sqrt{\mathcal{F}_{AB_I}}\right)}, \\
\mathcal{B}_{AB_{II}} &= \sqrt{2\left(1 - \sqrt{\mathcal{F}_{AB_{II}}}\right)}.
\end{align}
These expressions reveal the impact of Hawking radiation on the proximity between the evolving quantum state and its original entangled form, thereby offering an alternative and geometrically grounded characterization of quantum correlations.

\section{Results and Discussion \label{sec4}}  
This section presents a model designed to investigate the impact of gravity on quantum entanglement by considering a two‑qubit state in which one qubit interacts with a charged black hole. The Hawking effect transforms the initial bipartite state into a tripartite configuration, leading to a redistribution of quantum correlations. The black hole’s charge lowers its temperature, thereby reducing thermal effects. This framework allows for the analysis of information loss and the evolution of quantum resources in relativistic regimes.

\begin{figure}[H]
\centering
\begin{subfigure}[b]{0.45\textwidth}
    \includegraphics[width=\textwidth]{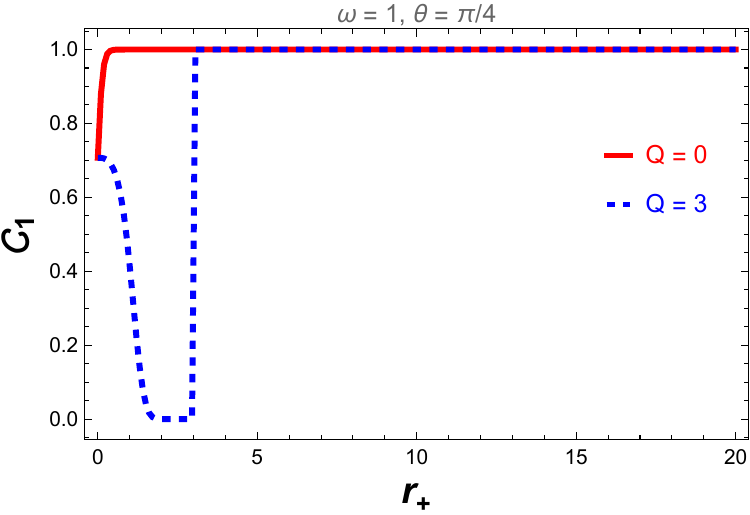}
    \caption{}
    \label{fig1a}
\end{subfigure}
\hfill
\begin{subfigure}[b]{0.45\textwidth}
    \includegraphics[width=\textwidth]{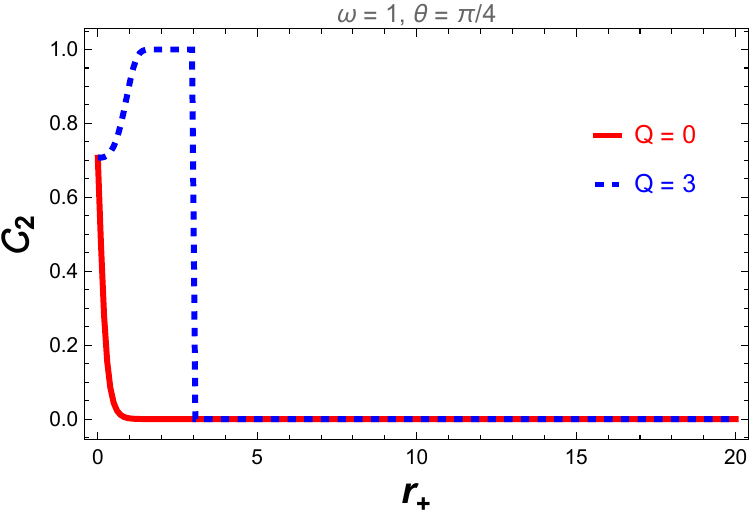}
    \caption{}
    \label{fig1b}
\end{subfigure}
\hfill
\begin{subfigure}[b]{0.45\textwidth}
    \includegraphics[width=\textwidth]{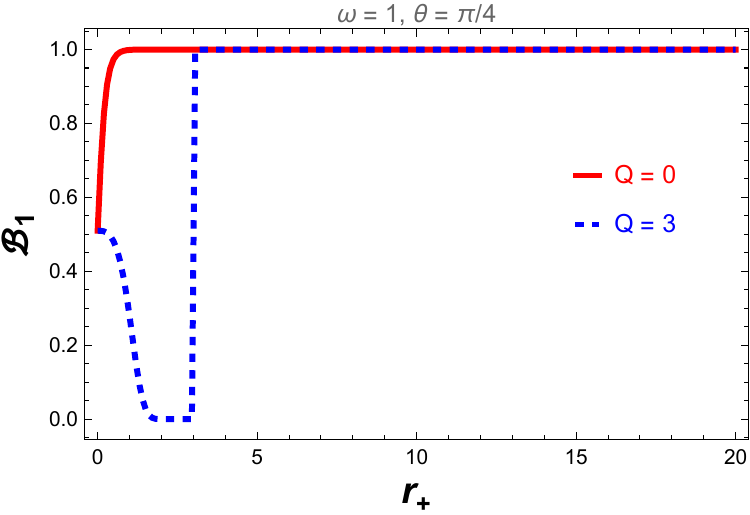}
    \caption{}
    \label{fig1c}
\end{subfigure}
\hfill
\begin{subfigure}[b]{0.45\textwidth}
    \includegraphics[width=\textwidth]{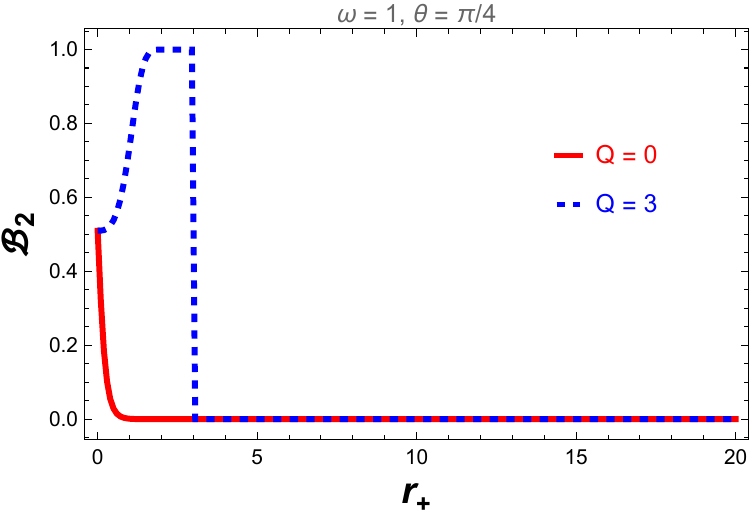}
    \caption{}
    \label{fig1d}
\end{subfigure}
\caption{Dynamics of $\mathcal{C}$ (\ref{fig1a}-\ref{fig1b}) and $\mathcal{B}$ (\ref{fig1c}-\ref{fig1d}) as functions the event horizon radius $r_+$ , in two scenarios: with and without charge, for observables 1 and 2
 with $\omega = 1$, $\theta = \pi/4 $, }
\label{figure1}
\end{figure}

Figure~\ref{figure1} illustrates the evolution of two quantum entanglement measures — the concurrence (\( \mathcal{C} \)) and the Bures distance (\( \mathcal{B} \)) — as functions of the event horizon radius \( r_+ \), within the framework of quantum field theory in curved spacetime. The analysis is conducted for a Reissner–Nordström black hole, considering both the uncharged case (\( Q = 0 \)) and the charged case (\( Q = 3 \)). These results provide insight into how Hawking radiation affects the initial quantum correlations between two subsystems.

The quantities \( \mathcal{C}_1 \) and \( \mathcal{B}_1 \) correspond to entanglement in Region I, located outside the event horizon where information remains physically accessible. In the absence of electric charge (\( Q = 0 \)), both measures remain high for small values of \( r_+ \), indicating that quantum correlations are preserved despite the gravitational background. However, when the electric charge is introduced (\( Q = 3 \)), a sharp decline in entanglement is observed beyond a certain horizon radius. This phenomenon is consistent with previous findings~\cite{bruschi2010unruh}, showing that the presence of an event horizon (due to gravity or acceleration) tends to degrade accessible quantum resources. Moreover, the electromagnetic field amplifies this degradation by modifying the vacuum structure ~\cite{fuentes2005alice}.

The quantities \( \mathcal{C}_2 \) and \( \mathcal{B}_2 \) represent the entanglement observed in Region II, which lies inside the horizon and is inaccessible to external observers. For the charged case (\( Q = 3 \)), a remarkable behavior emerges: over a narrow range of \( r_+ \), entanglement can be higher than in the uncharged case. This suggests that the electromagnetic field may temporarily enhance or stabilize certain components of entanglement. Such an effect is consistent with the tripartite distribution of Hawking modes described in~\cite{adesso2007multipartite}.

The observed asymmetry between Region I and Region II reflects the multipartite nature of the quantum state after Hawking emission. A portion of the initial entanglement is redirected to inaccessible modes across the horizon, leading to a perceived loss of entanglement from the perspective of an external observer. This behavior can be understood through the mechanism of gravitational decoherence, where quantum correlations are not destroyed but redistributed across causally disconnected regions~\cite{ver2009entangling,chandran2020one,ladghami2024barrow}.

\begin{figure}[H]
\centering
\begin{subfigure}[b]{0.45\textwidth}
    \includegraphics[width=\textwidth]{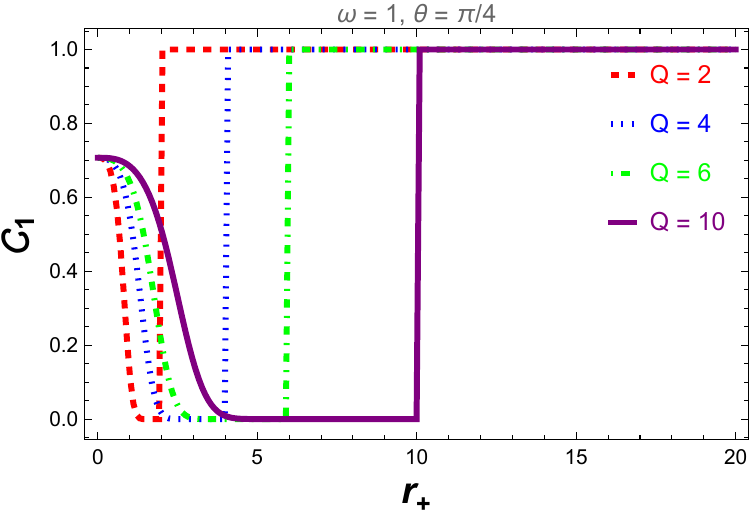}
    \caption{}
    \label{fig2a}
\end{subfigure}
\hfill
\begin{subfigure}[b]{0.45\textwidth}
    \includegraphics[width=\textwidth]{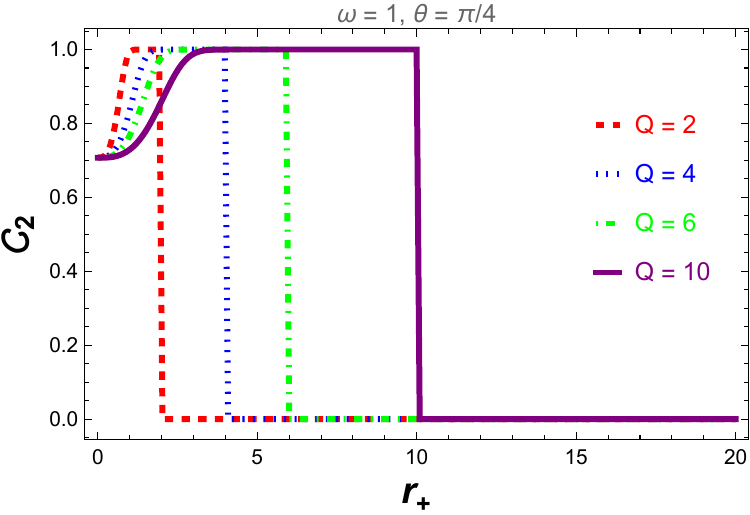}
    \caption{}
    \label{fig2b}
\end{subfigure}
\hfill
\begin{subfigure}[b]{0.45\textwidth}
    \includegraphics[width=\textwidth]{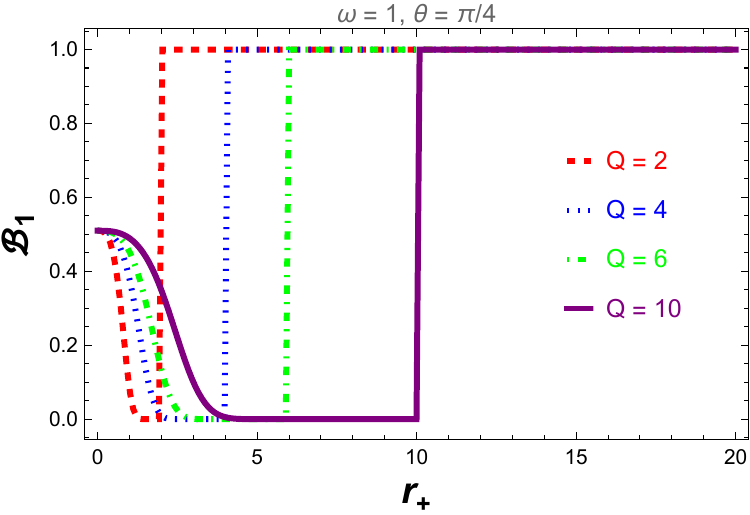}
    \caption{}
    \label{fig2c}
\end{subfigure}
\hfill
\begin{subfigure}[b]{0.45\textwidth}
    \includegraphics[width=\textwidth]{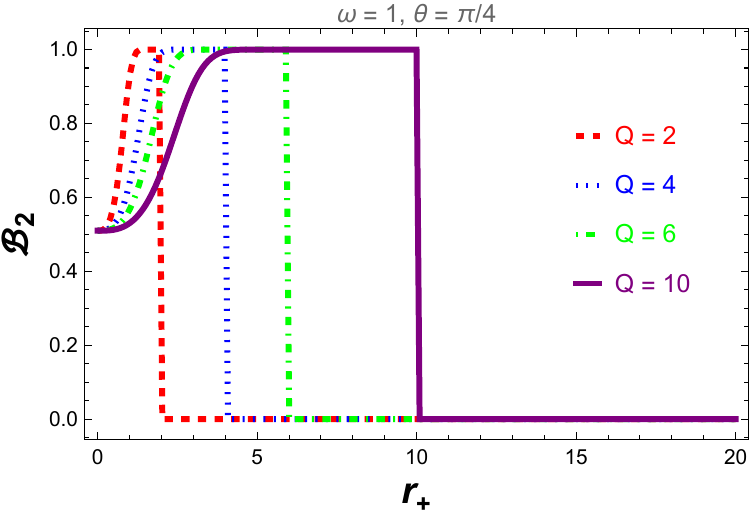}
    \caption{}
    \label{fig2d}
\end{subfigure}
\caption{Plots illustrating {C} (\ref{fig2a}-\ref{fig2b}),  
$B$ (\ref{fig2c}-\ref{fig2d}), as functions of the event horizon radius $r_+$ for various values of the black hole charge $Q$, with  
$ \theta = \pi/4$ and $\omega = 1$ , }

\label{figure2}
\end{figure}
Figure~\ref{figure2} illustrates the impact of the black hole’s electric charge \( Q \) on two fundamental quantum entanglement measures: the concurrence \( C \) and the Bures distance \( \mathcal{B} \), both plotted as functions of the event horizon radius \( r_+ \). The analysis considers fermionic field modes with frequency \( \omega = 1 \) and angular parameter \( \theta = \pi/4 \), with separate evaluations inside (region II) and outside (region I) the event horizon.

In the interior region, the curves for \( C_1 \) and \( \mathcal{B}_1 \) initially start at moderate values around 0.5 for small \( r_+ \lesssim 1 \), followed by a slight dip. However, beyond \( r_+ \approx 2 \), both measures exhibit a rapid rise to the maximal value of 1. This transition occurs at different thresholds depending on the charge: around \( r_+ \approx 2.8 \) for \( Q = 2 \), \( r_+ \approx 4.2 \) for \( Q = 4 \), \( r_+ \approx 6.5 \) for \( Q = 6 \), and \( r_+ \approx 9.5 \) for \( Q = 10 \). The results indicate a process of entanglement regeneration inside the black hole, despite extreme spacetime curvature. Such behavior can be interpreted in terms of a nonlocal redistribution of quantum field modes, consistent with the quantum field theory of particle creation in curved spacetime~\cite{hawking1975particle} and models of gravitational decoherence~\cite{blencowe2013effective}.
In the exterior region, the behavior of \( C_2 \) and \( \mathcal{B}_2 \) is also charge-sensitive. Both measures rise rapidly to values near 1 at intermediate \( r_+ \), then abruptly collapse as \( r_+ \) increases further. The extent of the entangled region grows with increasing \( Q \): for instance, entanglement disappears around \( r_+ \approx 4 \) for \( Q = 2 \), but persists up to \( r_+ \approx 9.5 \) for \( Q = 10 \). This suggests that the electric field plays a role in temporarily stabilizing quantum correlations against Hawking-induced decoherence. These observations are aligned with earlier studies on relativistic entanglement degradation and survival~\cite{fuentes2005alice, martin2011fermionic}.

Overall, the results reveal a marked asymmetry in entanglement behavior across the event horizon, regulated by the black hole charge \( Q \). Higher values of \( Q \) not only delay the loss of entanglement, but may also enhance or prolong it within specific spatial intervals. This supports the idea of a multipartite redistribution of entanglement across causally disconnected regions, a phenomenon central to quantum field theory in curved spacetime~\cite{birrell1984quantum}.

\begin{figure}[H]
\centering
\begin{subfigure}[b]{0.45\textwidth}
    \includegraphics[width=\textwidth]{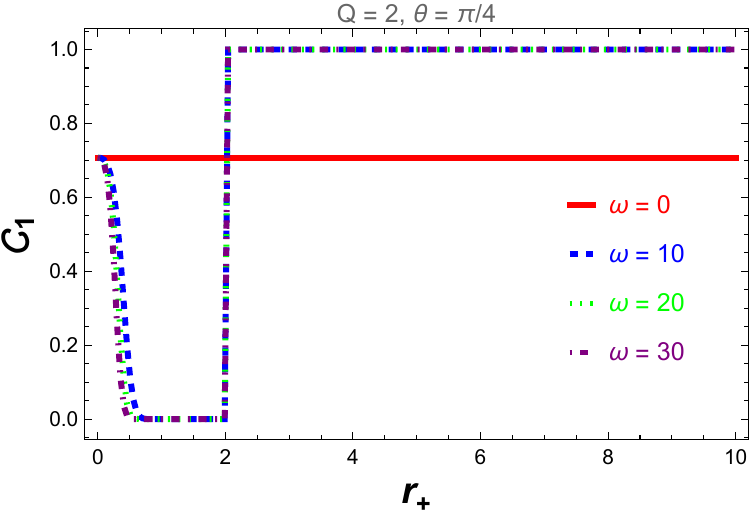}
    \caption{}
    \label{fig3a}
\end{subfigure}
\hfill
\begin{subfigure}[b]{0.45\textwidth}
    \includegraphics[width=\textwidth]{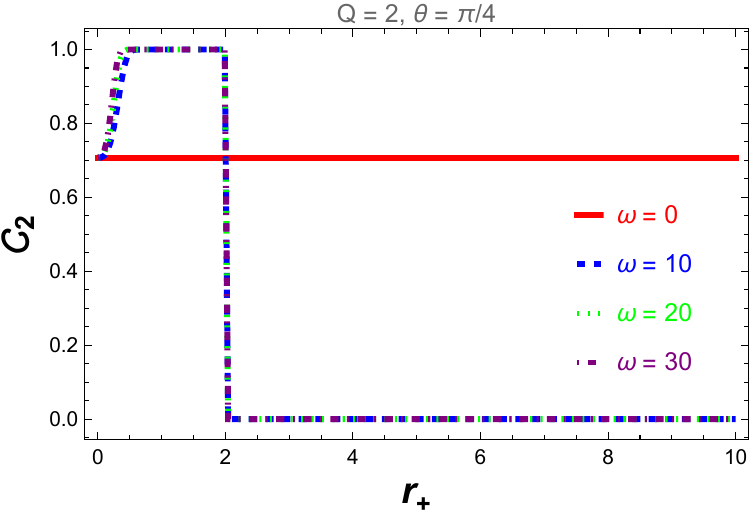}
    \caption{}
    \label{fig3b}
\end{subfigure}
\hfill
\begin{subfigure}[b]{0.45\textwidth}
    \includegraphics[width=\textwidth]{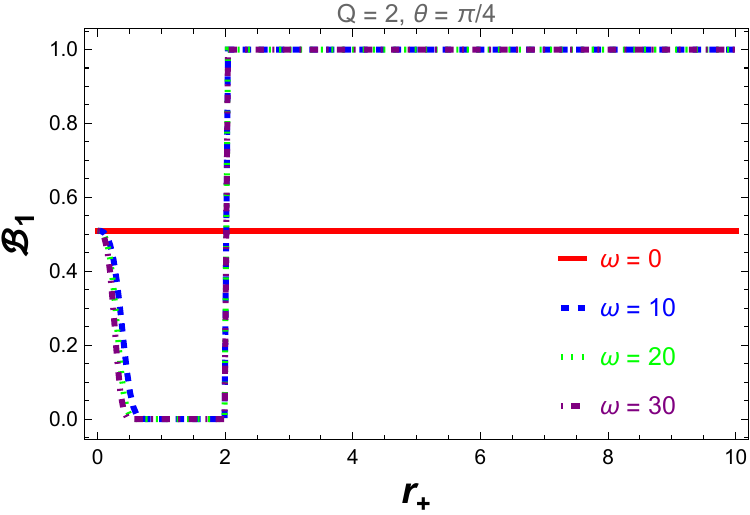}
    \caption{}
    \label{fig3c}
\end{subfigure}
\hfill
\begin{subfigure}[b]{0.45\textwidth}
    \includegraphics[width=\textwidth]{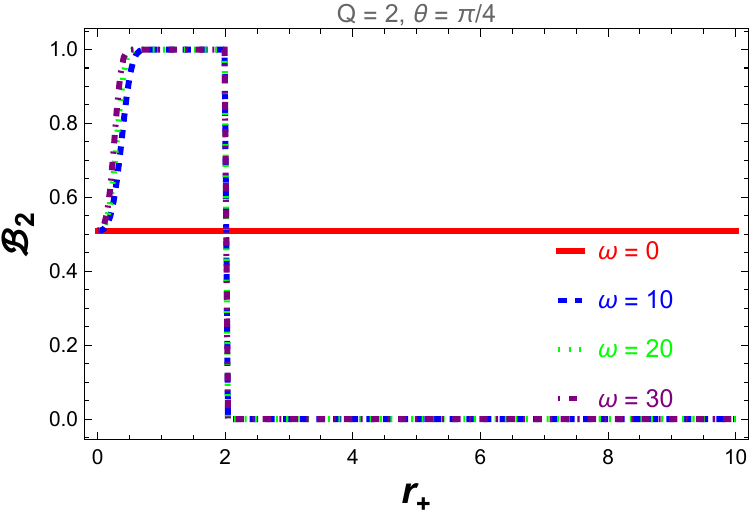}
    \caption{}
    \label{fig3d}
\end{subfigure}

\caption{Plots illustrating {C} (\ref{fig3a}-\ref{fig3b}),  
$B$ (\ref{fig3c}-\ref{fig3d}), as functions of the event horizon radius $r_+$ for various values of the frequency \( \omega \), with  
$  \theta = \pi/4$ and $Q = 2$, }
\label{figure3}
\end{figure}

Figure~\ref{figure3} analyzes how the frequency of fermionic modes, denoted by \( \omega \), influences the distribution and degradation of quantum entanglement near a Reissner–Nordström black hole. For a fixed charge \( Q = 2 \) and angular parameter \( \theta = \pi/4 \), we investigate the concurrence \( C \) and the Bures distance \( \mathcal{B} \) as functions of the event horizon radius \( r_+ \), considering four distinct frequencies: \( \omega = 0, 10, 20, 30 \). Two observers are considered: one inside the event horizon (region II) and one outside (region I), respectively shown in subfigures (\ref{fig3a}, \ref{fig3c}) and (\ref{fig3b}, \ref{fig3d}).

In the interior region, the entanglement behavior exhibits a strong dependence on the mode frequency. For low-frequency modes (\(\omega = 0\)), both the concurrence \(C_1\) and the Bures distance \(B_1\) remain nearly constant around moderate values (approximately \(0.6\)), despite variations in \(r_+\). However, as the frequency increases, a more pronounced pattern emerges: for \(\omega = 30\), both entanglement measures reach their maximum value (\(1\)) and remain stable over a wide interval (\(r_+ \lesssim 2\)), before abruptly dropping to zero. This indicates that high-frequency modes sustain stronger correlations under extreme gravitational conditions and exhibit greater resilience to decoherence within the horizon.

In the exterior region, the dependence on frequency is also significant, but the behavior is inverted. At low frequencies, the accessible entanglement—quantified by \(C_2\) and \(B_2\)—remains limited and shows minimal variation (e.g., \(C_2 \approx 0.6\), \(B_2 \approx 0.45\)).

As the frequency increases, entanglement becomes more pronounced: for \( \omega = 30 \), both measures rapidly increase to values near unity and persist up to \( r_+ \approx 2.5 \), after which a sudden drop occurs. This behavior reflects a relocalization of quantum correlations toward the exterior region for high-energy modes, facilitated by Bogoliubov mixing and the structure of the Hawking vacuum.

These results suggest that the mode frequency \( \omega \) acts as a control parameter for the accessibility and robustness of quantum entanglement in curved spacetime. While low-frequency modes remain more confined and entangled within the black hole interior, high-frequency excitations allow for stronger correlations to emerge outside. This frequency-dependent redistribution of entanglement highlights the nontrivial dynamics of quantum information across causal boundaries and supports the view of entanglement as a multipartite and frame-dependent resource in the presence of gravity~\cite{elghaayda2024physically}.

\begin{figure}[H]
\centering
\begin{subfigure}[b]{0.45\textwidth}
    \includegraphics[width=\textwidth]{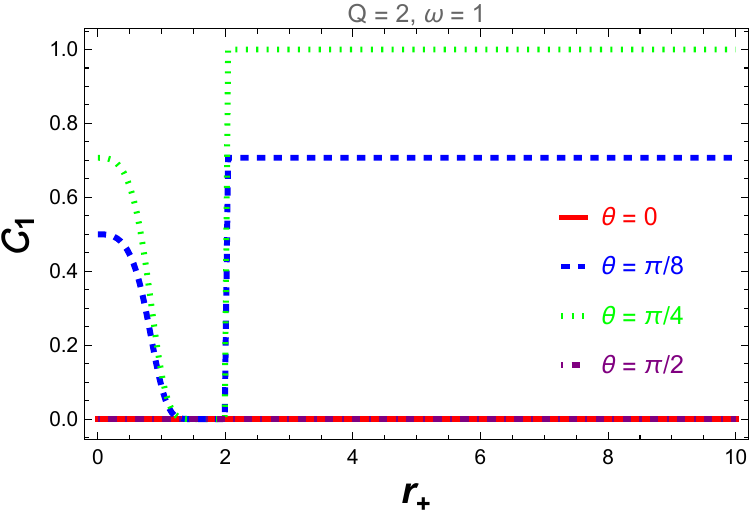}
    \caption{}
    \label{fig4a}
\end{subfigure}
\hfill
\begin{subfigure}[b]{0.45\textwidth}
    \includegraphics[width=\textwidth]{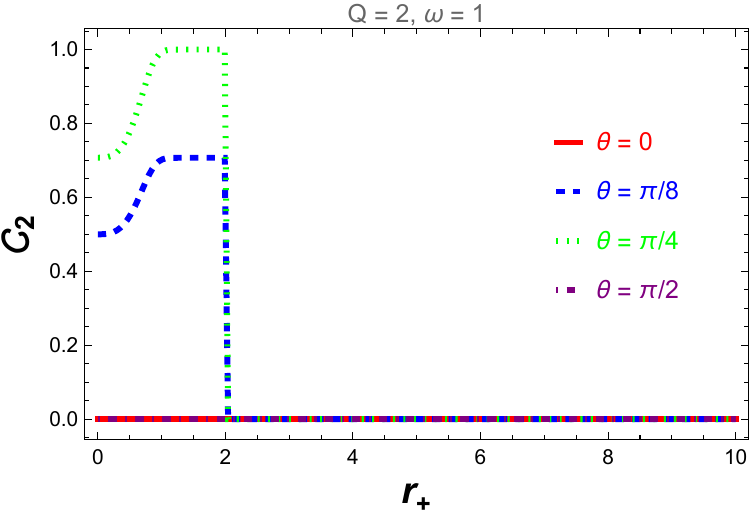}
    \caption{}
    \label{fig4b}
\end{subfigure}
\hfill
\begin{subfigure}[b]{0.45\textwidth}
    \includegraphics[width=\textwidth]{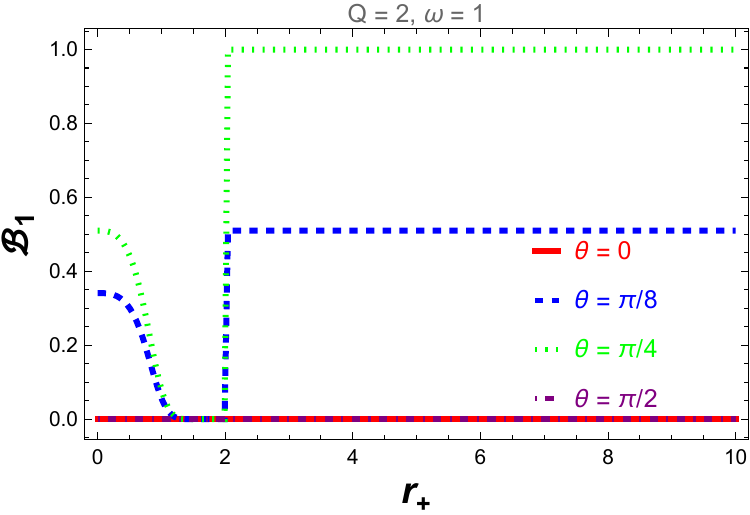}
    \caption{}
    \label{fig4c}
\end{subfigure}
\hfill
\begin{subfigure}[b]{0.45\textwidth}
    \includegraphics[width=\textwidth]{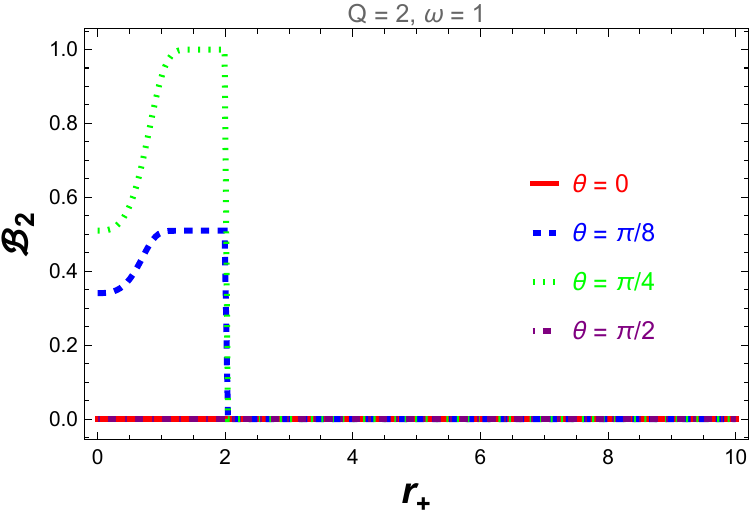}
    \caption{}
    \label{fig4d}
\end{subfigure}

\caption{Plots illustrating {C} (\ref{fig4a}-\ref{fig4b}),  
$B$ (\ref{fig4c}-\ref{fig4d}), as functions of the event horizon radius $r_+$ for various values of the angle \( \theta \), with  
$ \omega = 1$ and $Q = 2$, }
\label{figure4}
\end{figure}

Figure \ref{figure4} investigates how the initially entangled state $ \vert \psi \rangle_{AB} = \cos \theta \, \vert 00 \rangle + \sin \theta \, \vert 11 \rangle,$
influences the dynamics of quantum entanglement in the vicinity of a Reissner--Nordstr\"om black hole. Quantum resources, quantified via concurrence and the Bures distance, are analyzed for various values of the angle $\theta$, which controls the initial degree of entanglement.
The results demonstrate that when the system is prepared in a highly entangled state (particularly at $\theta = \pi/4$, corresponding to a Bell state), the entanglement exhibits greater robustness against Hawking radiation and spacetime deformation. In contrast, for $\theta = 0$ (state $\vert 00 \rangle$) or $\theta = \pi/2$ (state $\vert 11 \rangle$), the entanglement is weak or absent, and quantum resources decay rapidly.
These findings emphasize that the initial superposition structure plays a crucial role in preserving quantum correlations, consistent with previous studies by Fuentes-Schuller \textit{et al.} \cite{fuentes2005alice}, Alsing and Milburn \cite{Alsing2003}, and Martín-Martínez \textit{et al.} \cite{Martin-Martinez2010}. In summary, the initial state $\vert \psi \rangle_{AB}$ strongly conditions the ability of entanglement to survive under extreme gravitational conditions.

\begin{figure}[H]
\centering
\begin{subfigure}[b]{0.45\textwidth}
    \includegraphics[width=\textwidth]{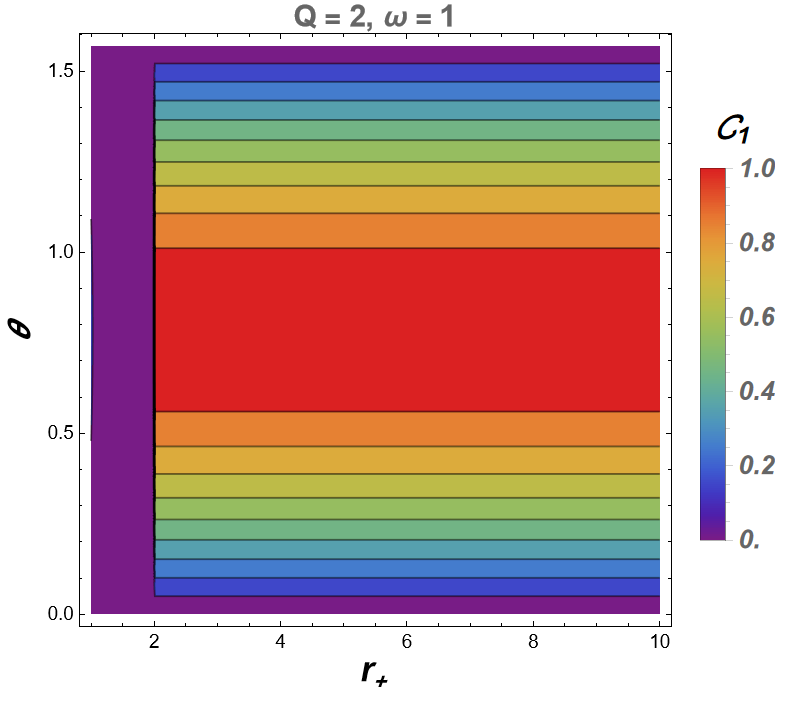}
    \caption{}
    \label{fig5a}
\end{subfigure}
\hfill
\begin{subfigure}[b]{0.45\textwidth}
    \includegraphics[width=\textwidth]{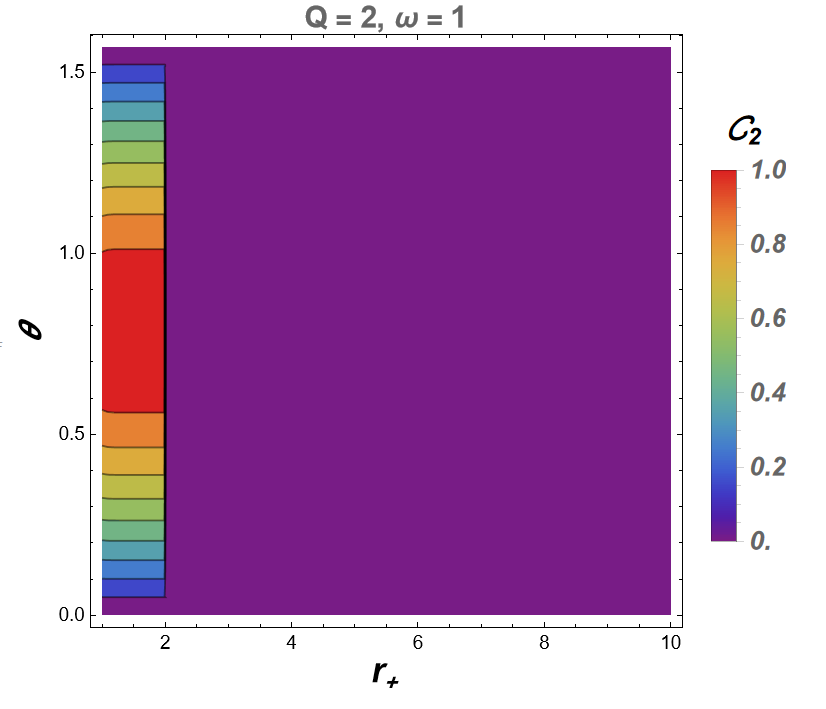}
    \caption{}
    \label{fig5b}
\end{subfigure}
\hfill
\begin{subfigure}[b]{0.45\textwidth}
    \includegraphics[width=\textwidth]{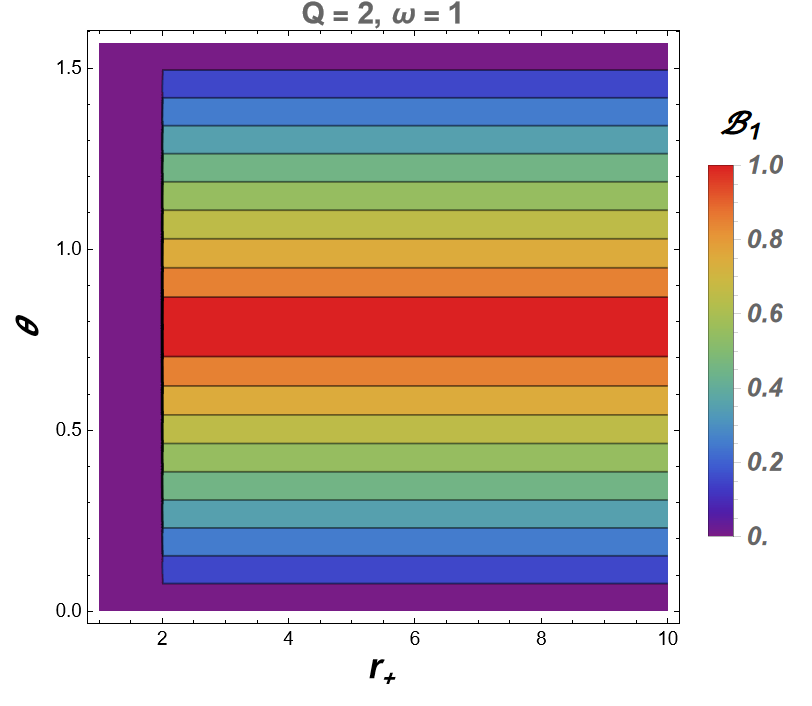}
    \caption{}
    \label{fig5c}
\end{subfigure}
\hfill
\begin{subfigure}[b]{0.45\textwidth}
    \includegraphics[width=\textwidth]{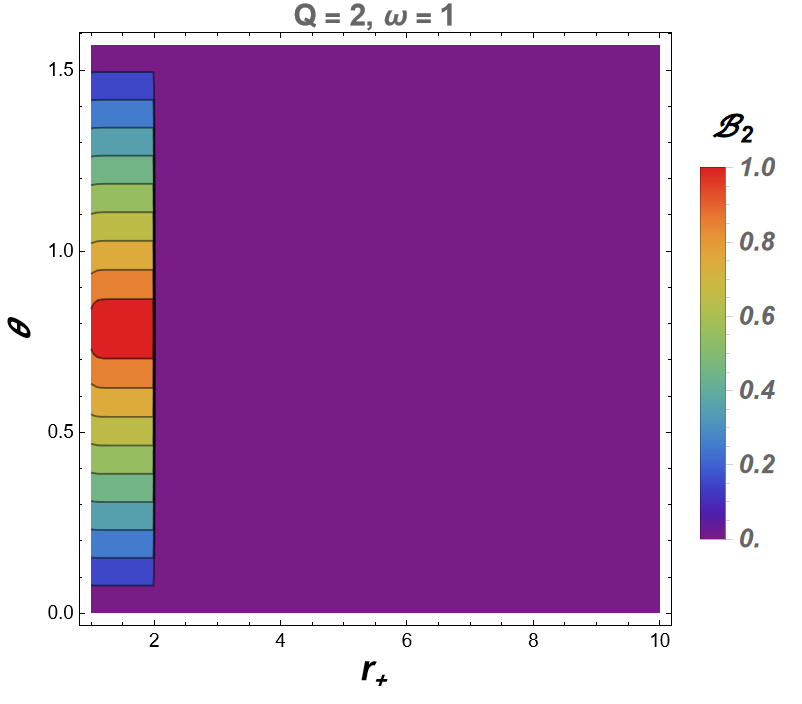}
    \caption{}
    \label{fig5d}
\end{subfigure}

\caption{Plots illustrating {C} (\ref{fig5a}-\ref{fig5b}),  
$B$ (\ref{fig5c}-\ref{fig5d}), as functions of the event horizon radius $r_+$ and $\theta$, with  
 \(\omega = 1\), $Q= 2$, }
\label{figure5}
\end{figure}

Figure \ref{figure5} illustrates the effect of the initial entanglement angle \( \theta \) on the evolution of quantum entanglement, measured by the concurrence \( C \) and the Bures distance \( \mathcal{B} \), as functions of the event horizon radius \( r_+ \), for a Reissner–Nordström black hole with \( Q = 2 \) and \( \omega = 1 \). Inside the black hole (region II), entanglement remains high and stable when the system is initially prepared in a strongly entangled state, especially for \( \theta = \pi/4 \), while it vanishes for separable states such as \( \theta = 0 \) or \( \theta = \pi/2 \). In contrast, in the exterior region (region I), the entanglement is significantly more localized and sensitive to the effects of Hawking radiation; it is only observable for small values of \( r_+ \) and intermediate angles. This behavior highlights the dynamic redistribution of quantum correlations in curved spacetime, leading to an apparent loss of entanglement for partial observers. It demonstrates that the robustness of entanglement under gravitational effects strongly depends on the initial configuration of the system, supporting the view that entanglement in curved spacetime is not destroyed but rather reorganized across causally disconnected regions.
\\
\section{Conclusion\label{sec5}}

In this work, we have analyzed the quantum entanglement of the Dirac field in the vicinity of Reissner–Nordström black hole, taking into account the effects of Hawking radiation within the framework of quantum field theory in curved spacetime~\cite{BirrellDavies, Hawking1975}. Using concurrence and Bures distance as entanglement quantifiers, we have highlighted how quantum correlations evolve depending on the black hole's electric charge \( Q \), the fermionic mode frequency \( \omega \), and the initial entanglement angle \( \theta \). Our results confirm that the presence of electric charge \( Q \) enhances decoherence within the event horizon, consistent with the findings of Martín-Martínez and Fuentes~\cite{Martin-Martinez2010}. Interestingly, however, it may also temporarily enhance the amount of accessible entanglement outside the horizon, in line with the multipartite redistribution of quantum correlations discussed in~\cite{bruschi2010unruh, VerSteeg2009}. The apparent loss of entanglement observed by an external observer results from a redistribution of correlations to causally disconnected regions, as supported by Bogoliubov transformations~\cite{Alsing2003}. Furthermore, our simulations reveal that high-frequency modes are more resilient to gravitational effects, maintaining robust correlations near the horizon, a finding in agreement with~\cite{Ahmadi2014}. This is also consistent with recent investigations into quantum steering and obesity in dilaton-like spacetimes~\cite{Elghaayda2024}. The initial entangled state plays a pivotal role in preserving correlations, with Bell-type states (\( \theta = \pi/4 \)) offering optimal robustness, as also reported in~\cite{FuentesSchuller2005, Alsing2006}.

These findings resonate with recent results on multipartite entanglement nonseparability~\cite{Wu2025_Nonseparability}, multiqubit coherence near black hole horizons~\cite{LiWu2025_Coherence}, and the metrological advantages of quantum correlations under relativistic effects~\cite{elghaayda2024physically}. Moreover, the behavior aligns with recent progress on entanglement harvesting across horizons~\cite{Wang2025_Harvesting}. 

This work opens the way to more in-depth studies on multipartite entanglement and complex quantum fields in dynamic spacetimes \cite{Henderson2020Multipartite}. Incorporating decoherence and environmental effects will allow assessing the robustness of quantum resources under realistic conditions \cite{Hu2021Decoherence}. These advances are essential for developing quantum protocols adapted to strong gravitational environments and could contribute to a better understanding of the black hole information paradox  \cite{Kole2019Relativistic}.

\printbibliography

@article{horodecki2009quantum,
  title     = {Quantum entanglement},
  author    = {Horodecki, Ryszard and Horodecki, Paweł and Horodecki, Michał and Horodecki, Karol},
  journal   = {Reviews of Modern Physics},
  volume    = {\textbf{81}},
  number    = {2},
  pages     = {865--942},
  year      = {2009},
  doi       = {10.1103/RevModPhys.81.865},
  publisher = {American Physical Society}
}

@article{brunner2014bell,
  title     = {Bell nonlocality},
  author    = {Brunner, Nicolas and Cavalcanti, Daniel and Pironio, Stefano and Scarani, Valerio and Wehner, Stephanie},
  journal   = {Reviews of Modern Physics},
  volume    = {\textbf{86}},
  number    = {2},
  pages     = {419--478},
  year      = {2014},
  doi       = {10.1103/RevModPhys.86.419},
  publisher = {American Physical Society}
}

@book{nielsen2010quantum,
  title     = {Quantum Computation and Quantum Information},
  author    = {Nielsen, Michael A. and Chuang, Isaac L.},
  edition   = {10th Anniversary Edition},
  publisher = {Cambridge University Press},
  year      = {2010},
  doi       = {10.1017/CBO9780511976667}
}

@article{fuentes2010entanglement,
  title     = {Entanglement in Curved Spacetimes},
  author    = {Fuentes, Ivette},
  journal   = {Classical and Quantum Gravity},
  volume    = {\textbf{31}},
  number    = {21},
  pages     = {214001},
  year      = {2014},
  doi       = {10.1088/0264-9381/31/21/214001},
  publisher = {IOP Publishing}
}

@book{birrell1984quantum,
  title     = {Quantum Fields in Curved Space},
  author    = {Birrell, N. D. and Davies, P. C. W.},
  publisher = {Cambridge University Press},
  year      = {1984},
  doi       = {https://doi.org/10.1017/CBO9780511622632}
}

@article{hawking1975particle,
  title     = {Particle Creation by Black Holes},
  author    = {Hawking, Stephen W.},
  journal   = {Communications in Mathematical Physics},
  volume    = {\textbf{43}},
  number    = {3},
  pages     = {199--220},
  year      = {1975},
  doi       = {10.1007/BF02345020},
  publisher = {Springer}
}

@article{AlsingMilburn2003,
  title     = {Teleportation with a Uniformly Accelerated Partner},
  author    = {Alsing, Paul M. and Milburn, G. J.},
  journal   = {Physical Review Letters},
  volume    = {\textbf{91}},
  pages     = {180404},
  year      = {2003},
  doi       = {10.1103/PhysRevLett.91.180404},
  publisher = {American Physical Society}
}

@article{FuentesSchuller2005,
  title     = {Alice Falls into a Black Hole: Entanglement in Noninertial Frames},
  author    = {Fuentes-Schuller, Ivette and Mann, Robert B.},
  journal   = {Physical Review Letters},
  volume    = {\textbf{95}},
  pages     = {120404},
  year      = {2005},
  doi       = {10.1103/PhysRevLett.95.120404},
  publisher = {American Physical Society}
}

@article{Montero2011,
  title     = {Fermionic Entanglement Ambiguity in Noninertial Frames},
  author    = {Montero, Marcelo and Martín-Martínez, Eduardo},
  journal   = {Physical Review A},
  volume    = {\textbf{83}},
  pages     = {062323},
  year      = {2011},
  doi       = {10.1103/PhysRevA.83.062323},
  publisher = {American Physical Society}
}

@article{neznamov2018stationary,
  title     = {Stationary Bound States of Spin‑Half Particles in the Reissner–Nordström Gravitational Field},
  author    = {Neznamov, V. P. and Safronov, I. I. and Shemarulin, V. E.},
  journal   = {Journal of Experimental and Theoretical Physics},
  volume    = {\textbf{127}},
  number    = {4},
  pages     = {684--704},
  year      = {2018},
  doi       = {10.1134/S1063776118100199},
  publisher = {Pleiades Publishing Ltd.}
}

@article{zhang2025entanglement,
  title     = {Entanglement and entropy uncertainty in black hole quantum atmosphere},
  author    = {Zhang, Shuai and Li, Li-Juan and Song, Xue-Ke and Ye, Liu and Wang, Dong},
  journal   = {Physics Letters B},
  volume    = {\textbf{868}},
  pages     = {139648},
  year      = {2025},
  doi       = {10.1016/j.physletb.2025.139648},
  publisher = {Elsevier}
}

@article{kanti2002calculable,
  title     = {Calculable corrections to brane black hole decay. II. Greybody factors for spin 1/2 and 1},
  author    = {Kanti, Panagiota and March-Russell, John},
  journal   = {Physical Review D},
  volume    = {\textbf{67}},
  number    = {10},
  pages     = {104019},
  year      = {2003},
  doi       = {10.1103/PhysRevD.67.104019},
  publisher = {American Physical Society}
}

@article{chhieb2024time,
  title     = {Time fractional evolution of two dipolar-coupled spins under DM and KSEA interactions},
  author    = {Chhieb, Abdessamie and Oumennana, Mansoura and Mansour, Mostafa and El Anouz, Khadija and Ouchrif, Mohamed},
  journal   = {Optical and Quantum Electronics},
  volume    = {\textbf{56}},
  number    = {9},
  pages     = {1421},
  year      = {2024},
  publisher = {Springer},
doi       = {https://doi.org/10.1007/s11082-024-07320-8}
}

@article{giovannetti2011advances,
  title     = {Advances in quantum metrology},
  author    = {Giovannetti, Vittorio and Lloyd, Seth and Maccone, Lorenzo},
  journal   = {Nature Photonics},
  volume    = {\textbf{5}},
  number    = {4},
  pages     = {222--229},
  year      = {2011},
  doi       = {10.1038/nphoton.2011.35},
  publisher = {Nature Publishing Group}
}

@article{uhlmann1976transition,
  title     = {The “transition probability” in the state space of a *$^*$-algebra},
  author    = {Uhlmann, Armin},
  journal   = {Reports on Mathematical Physics},
  volume    = {\textbf{9}},
  number    = {2},
  pages     = {273--279},
  year      = {1976},
  doi       = {10.1016/0034-4877(76)90060-4},
  publisher = {Elsevier}
}

@article{uhlmann1995geometric,
  title={Geometric phases and related structures},
  author={Uhlmann, Armin},
  journal={Reports on Mathematical Physics},
  volume={\textbf{36}},
  number={2-3},
  pages={461--481},
  year={1995},
  publisher={Pergamon},
doi       = {https://doi.org/10.1016/0034-4877(96)83640-8}

}

@article{fuentes2005alice,
  title     = {Alice Falls into a Black Hole: Entanglement in Noninertial Frames},
  author    = {Fuentes-Schuller, Ivette and Mann, Robert B.},
  journal   = {Physical Review Letters},
  volume    = {\textbf{95}},
  number    = {12},
  pages     = {120404},
  year      = {2005},
  doi       = {10.1103/PhysRevLett.95.120404},
  publisher = {American Physical Society}
}

@article{bruschi2010unruh,
  title     = {Unruh effect in quantum information beyond the single-mode approximation},
  author    = {Bruschi, David Edward and Louko, Jorma and Martín-Martínez, Eduardo and Dragan, Andrzej and Fuentes, Ivette},
  journal   = {Physical Review A},
  volume    = {\textbf{82}},
  number    = {4},
  pages     = {042332},
  year      = {2010},
  doi       = {10.1103/PhysRevA.82.042332},
  publisher = {American Physical Society}
}

@article{bennett1993teleporting,
  title     = {Teleporting an unknown quantum state via dual classical and Einstein-Podolsky-Rosen channels},
  author    = {Bennett, Charles H. and Brassard, Gilles and Crépeau, Claude and Jozsa, Richard and Peres, Asher and Wootters, William K.},
  journal   = {Physical Review Letters},
  volume    = {\textbf{70}},
  number    = {13},
  pages     = {1895--1899},
  year      = {1993},
  doi       = {10.1103/PhysRevLett.70.1895},
  publisher = {American Physical Society}
}

@book{chandrasekhar1983mathematical,
  title     = {The Mathematical Theory of Black Holes},
  author    = {Chandrasekhar, Subrahmanyan},
  publisher = {Oxford University Press},
  year      = {1983},
  isbn      = {9780198503705},
  address   = {Oxford, UK},
doi       = {https://doi.org/10.1007/978-94-009-6469-3_2}}

@article{chhieb2024metrological,
  title={Metrological non-classical correlations and quantum coherence in hybrid $(1/2, 1) $ system under decoherence channels},
  author={Chhieb, Abdessamie and Oumennana, Mansoura and Bouafia, Zakaria and Chouiba, Aicha and Mansour, Mostafa and Ouchrif, Mohamed},
  journal={Laser Physics},
  volume={\textbf{34}},
  number={10},
  pages={105202},
  year={2024},
doi       = {10.1088/1555-6611/ad71b0},
  publisher={IOP Publishing}
}

@article{banouni2025unveiling,
  title={Unveiling geometric quantum resources and uncertainty relation in a two-dimensional electron gas},
  author={Banouni, Chaimae and Bouafia, Zakaria and Mansour, Mostafa and Ouchrif, Mohamed},
  journal={Applied Physics B},
  volume={\textbf{131}},
  number={1},
  pages={9},
  year={2025},
doi       = {https://doi.org/10.1007/s00340-024-08368-w},
  publisher={Springer}
}

@article{alsing2003teleportation,
  title     = {Teleportation with a uniformly accelerated partner},
  author    = {Alsing, Paul M. and Milburn, G. J.},
  journal   = {Physical Review Letters},
  volume    = {\textbf{91}},
  number    = {18},
  pages     = {180404},
  year      = {2003},
  doi       = {10.1103/PhysRevLett.91.180404},
  publisher = {American Physical Society}
}

@article{mann2012relativistic,
  title     = {Relativistic quantum information},
  author    = {Mann, Robert B. and Ralph, Timothy C.},
  journal   = {Classical and Quantum Gravity},
  volume    = {\textbf{29}},
  number    = {22},
  pages     = {220301},
  year      = {2012},
  doi       = {10.1088/0264-9381/29/22/220301},
  publisher = {IOP Publishing}
}

@article{Carter1966,
  title     = {Complete Analytic Extension of the Symmetry Axis of Kerr's Solution of Einstein's Equations},
  author    = {Carter, Brandon},
  journal   = {Physical Review},
  volume    = {\textbf{141}},
  number    = {4},
  pages     = {1242--1247},
  year      = {1966},
  doi       = {10.1103/PhysRev.141.1242},
  publisher = {American Physical Society}
}

@article{Bardeen1973,
  title     = {The Four Laws of Black Hole Mechanics},
  author    = {Bardeen, James M. and Carter, Brandon and Hawking, Stephen W.},
  journal   = {Communications in Mathematical Physics},
  volume    = {\textbf{31}},
  number    = {2},
  pages     = {161--170},
  year      = {1973},
  doi       = {10.1007/BF01645742},
  publisher = {Springer}
}

@article{Jing2004,
  author  = {Felix Finster and Joel Smoller and Shing-Tung Yau},
  title   = {Nonexistence of time-periodic solutions of the Dirac equation in a Reissner–Nordström black hole background},
  journal = {Journal of Mathematical Physics},
  volume  = {\textbf{41}},
  number  = {4},
  pages   = {2173--2194},
  year    = {2000},
  doi     = {https://doi.org/10.1063/1.533234}
}

@article{Hawking1975,
  title     = {Particle Creation by Black Holes},
  author    = {Hawking, Stephen W.},
  journal   = {Communications in Mathematical Physics},
  volume    = {\textbf{43}},
  number    = {3},
  pages     = {199--220},
  year      = {1975},
  doi       = {10.1007/BF02345020},
  publisher = {Springer}
}

@article{HawkingPage1983,
  title     = {Thermodynamics of Black Holes in anti-De Sitter Space},
  author    = {Hawking, Stephen W. and Page, Don N.},
  journal   = {Communications in Mathematical Physics},
  volume    = {\textbf{87}},
  number    = {4},
  pages     = {577--588},
  year      = {1983},
  doi       = {10.1007/BF01208266},
  publisher = {Springer}
}

@article{wootters1998quantum,
  title     = {Entanglement of Formation of an Arbitrary State of Two Qubits},
  author    = {Wootters, William K.},
  journal   = {Physical Review Letters},
  volume    = {\textbf{80}},
  number    = {10},
  pages     = {2245--2248},
  year      = {1998},
  doi       = {10.1103/PhysRevLett.80.2245},
  publisher = {American Physical Society}
}

@article{adesso2007multipartite,
   title={Multipartite entanglement in three-mode Gaussian states of continuous-variable systems: Quantification, sharing structure, and decoherence},
  author={Adesso, Gerardo and Serafini, Alessio and Illuminati, Fabrizio},
  journal={Physical Review A—Atomic, Molecular, and Optical Physics},
  volume={\textbf{73}},
  number={3},
  pages={032345},
  year={2006},
  publisher={APS},
  doi={https://doi.org/10.1103/PhysRevA.73.032345}
}

@article{blencowe2013effective,
  title={Effective Field Theory Approach to Gravitationally Induced Decoherence},
  author={Blencowe, M. P.},
  journal={Physical Review Letters},
  volume={\textbf{111}},
  number={2},
  pages={021302},
  year={2013},
  doi={10.1103/PhysRevLett.111.021302}
}

@article{martin2011fermionic,
  title={Fermionic entanglement ambiguity in noninertial frames},
  author={Montero, Miguel and Martin-Martinez, Eduardo},
  journal={Physical Review A},
  volume={\textbf{83}},
  number={6},
  pages={062323},
  year={2011},
  doi={10.1103/PhysRevA.83.062323}
}

@article{elghaayda2024physically,
  title={Physically Accessible and Inaccessible Quantum Correlations of Dirac Fields in Schwarzschild Spacetime},
  author={Elghaayda, Samira and Ali, Asad and Al‑Kuwari, Saif and Mansour, Mostafa},
  journal={Physics Letters A},
  volume={\textbf{525}},
  pages={29915},
  year={2024},
  doi={10.1016/j.physleta.2024.29915}
}

@article{Alsing2003,
  title={Teleportation with a uniformly accelerated partner},
  author={Alsing, Paul M. and Milburn, G. J.},
  journal={Physical Review Letters},
  volume={\textbf{91}},
  number={18},
  pages={180404},
  year={2003},
  doi={10.1103/PhysRevLett.91.180404}
}

@article{Martin-Martinez2010,
  title={Quantum entanglement produced in the formation of a black hole},
  author={Mart{\'\i}n-Mart{\'\i}nez, Eduardo and Garay, Luis J. and Le{\'o}n, Juan},
  journal={Physical Review D},
  volume={\textbf{82}},
  number={6},
  pages={064028},
  year={2010},
  doi={10.1103/PhysRevD.82.064028}
}

@book{BirrellDavies,
  title     = {Quantum Fields in Curved Space},
  author    = {Birrell, N. D. and Davies, P. C. W.},
  publisher = {Cambridge University Press},
  address   = {Cambridge},
  year      = {1982},
  isbn      = {978-0-521-23385-7},
  doi     = {10.1017/CBO9780511622632}
}

@article{Elghaayda2024,
  title={Quantum obesity and steering ellipsoids for fermionic fields in Garfinkle-Horowitz-Strominger dilation spacetime},
  author={Elghaayda, Samira and Abd-Rabbou, MY and Mansour, Mostafa},
  journal={arXiv preprint arXiv:2408.06869},
  year={2024},
  doi={https://doi.org/10.48550/arXiv.2408.06869}
}

@article{Alsing2006,
  title={Entanglement of Dirac fields in noninertial frames},
  author={Alsing, Paul M. and Fuentes‑Schuller, Ivette and Mann, Robert B. and Tessier, Terrance E.},
  journal={Physical Review A},
  volume={\textbf{74}},
  number={3},
  pages={032326},
  year={2006},
  doi={10.1103/PhysRevA.74.032326}
}

@article{VerSteeg2009,
  title   = {Entangling power of an expanding universe},
  author  = {Ver Steeg, Greg and Menicucci, Nicolas C},
  journal = {Physical Review D},
  volume  = {\textbf{79}},
  number  = {4},
  pages   = {044027},
  year    = {2009},
  doi     = {10.1103/PhysRevD.79.044027},
}

@article{Ahmadi2014,
  title={Relativistic quantum metrology: Exploiting relativity to improve quantum measurement technologies},
  author={Ahmadi, Mehdi and Bruschi, David E. and Sab{\'\i}n, Carlos and Adesso, Gerardo and Fuentes, Ivette},
  journal={Scientific Reports},
  volume={\textbf{4}},
  pages={4996},
  year={2014},
  doi={10.1038/srep04996}
}

@article{Wu2025_Nonseparability,
  title={Nonseparability of multipartite systems in dilaton black hole},
  author={Wu, Shu-Min and Teng, Xiao-Wei and Li, Wen-Mei and Wang, Yu-Xuan and Lu, Jianbo},
  journal={arXiv preprint arXiv:2503.17923},
  year={2025},
 doi={https://doi.org/10.48550/arXiv.2503.17923}
}

@article{LiWu2025_Coherence,
  title={Duality relation between coherence and path information in the presence of quantum memory},
  author={Li, J. and Wu, L. and Fei, S. M.},
  journal={Journal of Physics A: Mathematical and Theoretical},
  volume={\textbf{51}},
  number={8},
  pages={085304},
  year={2018},
  doi={10.1088/1751-8121/aaa0f6}
}

@article{Wang2025_Harvesting,
  title={Harvesting Information Across the Horizon},
  author={Wang, S. and Preciado Rivas, M. R. and Mann, R. B.},
  journal={arXiv},
  year={2025},
  eprint={2504.00083},
 doi={https://doi.org/10.48550/arXiv.2504.00083}
}

@article{ver2009entangling,
  title={Entangling power of an expanding universe},
  author={Ver Steeg, Greg and Menicucci, Nicolas C.},
  journal={Physical Review D},
  volume={\textbf{79}},
  number={4},
  pages={044027},
  year={2009},
  doi={10.1103/PhysRevD.79.044027}
}

@article{chandran2020one,
  title={One-to-one correspondence between entanglement mechanics and black hole thermodynamics},
  author={Chandran, S. Mahesh and Shankaranarayanan, S.},
  journal={Physical Review D},
  volume={\textbf{102}},
  pages={125025},
  year={2020},
  doi={10.1103/PhysRevD.102.125025}
}

@article{ladghami2024barrow,
  title={Barrow entropy and AdS black holes in RPS thermodynamics},
  author={Ladghami, Yahya and others},
  journal={Physics of the Dark Universe},
  volume={\textbf{44}},
  pages={101212},
  year={2024},
  doi={https://doi.org/10.1016/j.dark.2024.101470}
}

@article{Peres2004Quantum,
  author  = {Asher Peres and Daniel R. Terno},
  title   = {Quantum information and relativity theory},
  journal = {Reviews of Modern Physics},
  volume  = {\textbf{76}},
  number  = {1},
  pages   = {93--123},
  year    = {2004},
  doi     = {10.1103/RevModPhys.76.93}
}

@article{Rideout2012Fundamental,
  author  = {D. Rideout and others},
  title   = {Fundamental quantum optics experiments conceivable with satellites—reaching relativistic distances and velocities},
  journal = {Classical and Quantum Gravity},
  volume  = {\textbf{29}},
  number  = {22},
  pages   = {224011},
  year    = {2012},
  doi     = {10.1088/0264-9381/29/22/224011}
}

@article{fuentes2010diracEntanglement,
  author  = {Ivette Fuentes and Mercedes Martín-Martínez and Marcus B. Plenio and Enrique Solano},
  title   = {Entanglement of Dirac fields in an expanding spacetime},
  journal = {Physical Review D},
  volume  = {\textbf{82}},
  number  = {4},
  pages   = {045030},
  year    = {2010},
  doi     = {10.1103/PhysRevD.82.045030}
}

@article{Henderson2020Multipartite,
  author  = {L. Henderson and M. Vedral},
  title   = {Multipartite quantum correlations in relativistic settings},
  journal = {Physical Review A},
  volume  = {\textbf{101}},
  pages   = {012333},
  year    = {2020},
  doi     = {10.1103/PhysRevA.101.012333}
}

@article{Hu2021Decoherence,
  title={Decoherence due to spacetime curvature},
  author={Singh, Raghvendra and Khanna, Kabir and Kothawala, Dawood},
  journal={arXiv preprint arXiv:2302.09038},
  year={2023},
 doi     = {https://doi.org/10.48550/arXiv.2302.09038}
}

@article{Kole2019Relativistic,
  author  = {Daniel Harlow},
  title   = {Jerusalem lectures on black holes and quantum information},
  journal = {Reviews of Modern Physics},
  volume  = {\textbf{88}},
  pages   = {015002},
  year    = {2016},
  doi     = {10.1103/RevModPhys.88.015002}
}

\end{document}